\documentclass[apj]{emulateapj}
\usepackage{epsfig}
\usepackage{epstopdf}
\usepackage{graphics}

\begin{document}

\title[]{Pinpointing the knee of cosmic rays with diffuse PeV $\gamma$-rays and neutrinos}
\author{Y. Q. Guo, H. B. Hu, Q. Yuan, Z. Tian and X. J. Gao}

\affil{Key Laboratory of Particle Astrophysics, Institute of High 
Energy Physics, Chinese Academy of Science, Beijing 100049, PR China
}

\begin{abstract}
The origin of the knee in the cosmic ray spectrum remains to be an unsolved
fundamental problem. There are various kinds of models that predict
different break positions and the compositions of the knee. In this work, 
we suggest the use of diffuse $\gamma$-rays and neutrinos as probes to test 
these models. Based on several typical types of composition models, 
the diffuse $\gamma$-ray and neutrino spectra are calculated and 
show distinctive cutoff behaviors at energies from tens of TeV to multi-PeV.
The expected flux will be observable by the newly upgraded 
Tibet-AS$\gamma$+MD (muon detector) experiment as well as more sensitive 
future projects, such as LHAASO and HiSCORE. By comparing the neutrino
spectrum with the recent observations by the IceCube experiment, we find that 
the diffuse neutrinos from interactions between the cosmic rays and the
interstellar medium may not be responsible to the majority of the IceCube
events. Future measurements of the neutrinos may be able to identify the 
Galactic diffuse component and shed further light on the problem of the 
knee of cosmic rays.
\end{abstract}

\maketitle

\section{Introduction}

The knee of the cosmic ray (CR) spectrum was discovered more than 50 
yr ago \citep{KK1958}. However its underlying physical causes are still under 
debate \citep{2004APh....21..241H}. The most popular scenario is the 
so-called ``poly-gonato'' model, in which each composition of CRs has
its own knee and the superposition of all compositions form the observed
knee structure of the CR spectra \citep{2003APh....19..193H}. 
Phenomenologically the knee of each composition could be charge dependent
due to e.g. the acceleration limit or propagation leakage 
\citep{1983A&A...125..249L,1988ApJ...333L..65V,1993A&A...268..726P,
1996APh.....5..367B,1998A&A...330..389W}, 
mass dependent due to the interactions \citep{1993APh.....1..229K,
2001ICRC....5.1760K,2002APh....17...23C,2009ApJ...700L.170H,
2010SCPMA..53..842W,2013NJPh...15a3053G} or even constant 
\citep{2003APh....19..193H}. 
The three types of phenomenological models can all fit the all-particle
spectrum, however, the spectrum of each composition shows distinctive
behaviors \citep{2003APh....19..193H}. A precise measurement of the
spectrum of individual composition in PeV energies will be essential
for the understanding of the knee puzzle.

Apart from the direct measurement of the spectra of individual nuclei, 
diffuse $\gamma$-ray and neutrino spectra carry exact information on CR 
spectrum in energy-per-nucleon and thus are important in testing the 
composition models. The diffuse $\gamma$-ray emission with energy from 
muti-MeV to sub-TeV has been well studied by EGRET \citep{1997ApJ...481..205H} 
and $Fermi$-LAT \citep{2012ApJ...750....3A} experiments. Although the electron 
processes contribute a proper fraction of the observed flux, it has been 
well established that the diffuse $\gamma$-ray emission in the Galactic 
plane is dominated by the hadronic nuclei-nuclei collisions 
\citep{2000ApJ...537..763S,2004ApJ...613..962S}. Extending this scenario 
to the knee energy region, we expect a guaranteed source of the diffuse 
sub-PeV $\gamma$-rays and neutrinos from the collision of the CRs and 
the interstellar medium (ISM) in the Galactic plane. 
Given the fact that
electrons suffer strong energy losses at PeV energies, the diffuse
$\gamma$-rays and neutrinos should be dominantly related with the
hadronic process.

Very-high-energy diffuse $\gamma$-rays have been extensively explored 
by ground-based extensive air shower (EAS) array experiments. In the PeV 
energy region, an upper limit on $\gamma$-ray flux was reported by 
CASAMIA \citep{1997PhRvL..79.1805C, 1998ApJ...493..175B}, 
KASCADE \citep{2003ICRC....4.2293S}, EAS-TOP \citep{1992ApJ...397..148A}, 
UMC \citep{1991ApJ...375..202M} and IC40 \citep{2013PhRvD..87f2002A}. In 
the muti-TeV region, HEGRA \citep{1995PhLB..347..161K}, Tibet-AS$\gamma$
\citep{2002ApJ...580..887A} reported flux upper limit, while MILAGRO 
\citep{2008ApJ...688.1078A} and ARGO-YBJ \citep{2011ICRC....7....2M} gave 
measurements in limited sky regions of the Galactic plane. 

Recently the very-high-energy neutrino observation has made great 
progress thanks to the IceCube experiment. The IceCube collaboration 
reported the detections of two PeV neutrino events and 26 other 
neutrino events from 20 to 400 TeV \citep{2013PhRvL.111b1103A,2013Sci...342E...1I}. 
The number of events exceeds the standard atmospheric background estimate 
by 2.8$\sigma$ and 3.3$\sigma$ respectively, and may imply an 
astrophysical origin \citep{2013Sci...342E...1I}. Several works have discussed 
the Galactic contribution to these neutrino events, such as the TeV 
$\gamma$-ray sources \citep{2013ApJ...774...74F,2014PhRvD..89j3002N}, 
the Galactic center and Fermi bubbles \citep{2013PhRvD..88h1302R,
2013arXiv1309.4077A,2010ApJ...724.1044S}, and the diffuse component due 
to CR interaction with the ISM \citep{2013APh....48...75G,2014MNRAS.439.3414J}. 
For a recent review please see \citet{2014JHEAp...1....1A}. A general 
conclusion is that a hard neutrino spectrum $\sim E_{\nu}^{-2.0}$
with a cutoff at a few PeV, or a slightly softer one $E_{\nu}^{-2.3}$ for the 
single power-law assumption \citep{2014PhRvD..89h3003A} was inevitable in 
order to explain the PeV observation. 

In this work, we investigate the effect of the composition models on the 
diffuse $\gamma$-ray and neutrino spectra. Through comparing the calculated 
fluxes with expected sensitivity curves of existent or future experiments, 
the possibility to test the composition models by diffuse $\gamma$-ray
and neutrino measurements is discussed. In the next section, we briefly 
describe the phenomenological poly-gonato model \citep{2003APh....19..193H} 
used in the following calculation. In Section 3, we present the calculated  
$\gamma$-ray and neutrino spectra. Finally, Section 4 is the conclusion.

\section{Composition models in the knee region}

Although all particle spectrum has been well-measured in the energy 
region around the knee, precise measurements for individual species are 
not yet available. Limited by small effective area, space and balloon 
borne experiments can measure the CR spectra with energies less than 
a few hundred TeV. On the other hand, ground-based EAS experiments have 
large enough effective area but bear a rather large systematics to
distinguish different nuclei species. Therefore the model to describe
the knee is basically based on the extrapolation of low-energy measurements,
some physical considerations as well as the fit to the all particle spectrum.
\citep{2003APh....19..193H} summarized the typical three types of models
of the knee. We briefly describe them here.

The first type of the model is motivated by the diffusive shock acceleration 
(DSA) or propagation process. According to the DSA theory, CRs will have 
a power-law spectrum with a maximum energy cutoff. The cutoff energy depends 
on the source properties. For supernova remnants, the cutoff energy is 
estimated to be about PeV \citep{1983A&A...125..249L}. Simply because the 
acceleration energy is proportional to the charge of the accelerated 
particle, the cutoff energy for different species should be $Z$-dependent 
\citep{1983A&A...125..249L}. From the point of view of CR propagation, 
the knee structure might be a consequence of the leakage of CRs from the 
Galaxy \citep{1993A&A...268..726P}. As the gyromagnetic radius of a particle 
is proportional to the rigidity, the knee structure of the individual nuclei 
should also be $Z$-dependent. 

The second type of model is motivated by the interaction processes.
In these models, either threshold interactions or new physics at a certain 
energy scale can lead to a change in the measured flux or energy 
\citep{1993APh.....1..229K,2001ICRC....5.1760K,2002APh....17...23C,
2009ApJ...700L.170H}. As the threshold energy is related to the Lorentz 
factor of the CR particle, the break energy is expected to be $A$-dependent. 
The third type of the break is constant for all species. It is not well
physically motivated, but it might be a simple possibility.

We note that the problem of the knee was still open until now. There is no 
consensus about the origin of the knee, mainly due to the limited knowledge 
about the individual spectrum of each composition. Therefore our discussion 
is based on three typical kinds of phenomenological approaches to the knee, 
i.e., the poly-gonato model with $A$ or $Z$ dependent or constant break 
energies \citep{2003APh....19..193H}. It is possible that the problem may be 
even more complicated, and the results in the specific model may be different from 
these phenomenological approaches. We expect that these three approaches can
be typical representatives of various kinds of physical models. As an example,
we adopt the model incorporating photon-nuclei pair production 
\citep{2009ApJ...700L.170H,2010SCPMA..53..842W} to show the result of a 
physically based model. In this model, the knee of each composition is 
approximately $A$-dependent. However, the energy spectrum after the interactions 
is not simply a broken power law or a power law with cutoff. Due to the detailed 
interaction energy losses, there is both a pile-up effect and a high-energy tail of the 
particle spectrum \citep{2009ApJ...700L.170H,2010SCPMA..53..842W}.
It is worth noting that the KASCADE experiment recently reported a discovery
of ankle-like spectrum for light elements at about 10 PeV 
\citep{2005APh....24....1A}. 
   
Following \citep{2003APh....19..193H}, the CR spectrum of each composition 
is parameterized as
\begin{equation}
\frac {d\Phi_i}{dE}(E) = \Phi_i^0E^{-\gamma_i} \left [
1+\left (\frac{E}{\hat{E_i}}\right)^{\epsilon_c} \right ]
^{\frac{-\Delta \gamma}{\epsilon_c}}
\end{equation}
where $\Phi_i^0$ and $\gamma_i$ are the normalization and spectrum index
before the break of the $i$th species, $E$ is the total energy of the 
particle, $\Delta \gamma$ and $\epsilon_c$ characterize the change in
the spectrum at the break energy $\hat{E_i}$. The break energy 
$\hat{E_i}$ is expressed as
\begin{equation}
\hat{E_i} = \left \{
     \begin{array}{ll}
     \hat{E_p}\cdot Z &,\ \ \ {\rm charge~dependent}\\
     \hat{E_p}\cdot A &,\ \ \ {\rm mass~dependent}\\
     \hat{E_p}       &,\ \ \ {\rm constant}
     \end{array}
     \right .
\end{equation}
where $\hat{E_p}$ is the break energy of protons. The values of the
key parameters of the poly-gonato model are listed in Table 1 
\citep{2003APh....19..193H}.
\begin{table}[h]
\begin{center}
\caption{Parameters $\Delta \gamma$, $\epsilon_c$ and $\hat{E_p}$}
\begin{tabular}{cccc}\hline
   Model & $\hat{E_p}$(PeV) & $\Delta \gamma$ & $\epsilon_c$ \\
   $Z$-dependent break & 4.49 &  2.1  & 1.9 \\
   $A$-dependent break & 3.81 &  5.7  & 2.34 \\
   Constant break      & 3.68 &  0.44 & 1.84 \\\hline
\end{tabular}
\end{center}
\end{table}

As for the physically based interaction model, the flux of individual 
nuclei depends on the energy loss rate and can not be analytically
parameterized. We adopt the numerical results presented in 
\citet{2010SCPMA..53..842W} in the calculation.

\begin{figure}
\centering
\includegraphics[width=0.48\textwidth]{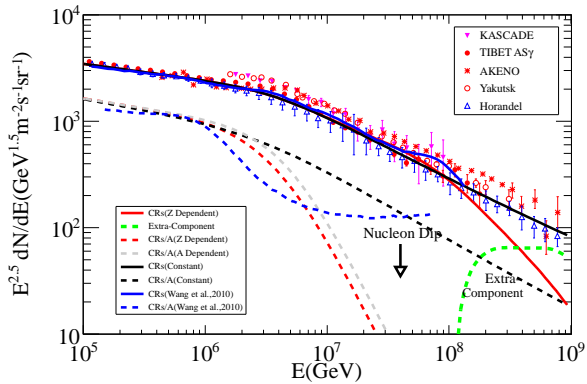}
\caption{Energy spectra of CRs. The solid lines are the all particle
spectra for $Z$-dependent break (red), constant break (black), and the 
physical interaction model (blue) respectively. The all particle spectrum 
of the $A$-dependent break model gives essentially the same result with the
$Z$-dependent break model, and is not shown. Dashed lines show the 
corresponding nucleon spectra. References of the data are: 
\citep{2005APh....24....1A,2001ICRC....1..145K,1992JPhG...18..423N,
2008ApJ...678.1165A,2003APh....19..193H}.}
\label{fig:crModel}
\end{figure}

The flux of all of the particle spectrum can be obtained by summing over 
all CR species
\begin{equation}
\frac{d\Phi}{dE}(E) = \sum\limits_{i}\frac{d\Phi_{i}}{dE}(E).
\end{equation}
The all particle spectra are shown by solid lines in Figure \ref{fig:crModel}. 
Note that an extra component might be necessary to explain the sub-EeV data
for $Z$- or $A$-dependent scenarios \citep{2005JPhG...31R..95H}.

The inelastic hadronic interactions are characterized by the nucleons in the 
particles. For the discussion of the $\gamma$-ray and neutrino production,
we convert the previous particle spectrum to nucleon spectrum
\begin{equation}
\begin{array}{lll}
\frac{d\Phi_n}{dE_n}&=&A\cdot\frac{d\Phi}{d(E/A)}  \\
   &=&A^{-\gamma_i+2}\Phi_i^0E_n^{-\gamma_i}\left[
   1+\left(\frac{E_n}{\hat{E_i}/A}\right)^{\epsilon_c}
   \right ]^{\frac{-\Delta\gamma}{\epsilon_c}}
\end{array}
\end{equation}
where $E_n=E/A$ is the energy per nucleon. We find that the nucleon spectrum
will be suppressed by a factor of $A^{-\gamma_i+2}$ compared with the all
particle spectrum, and the break energy is also different. As an example, 
for iron nucleus, the suppression factor is $56^{-2.59+2}=0.09$. 
The expected nucleon spectra of different models are also shown in 
Figure \ref{fig:crModel} by dashed lines. A very interesting feature of the 
spectrum is the existence of a ``dip'' structure for either charge-$Z$- 
or mass-$A$-dependent models. 

\section{Gamma-rays and neutrinos from CR interaction with ISM}

As we have discussed earlier, the dominant contribution of the diffuse
$\gamma$-rays and neutrinos with energy higher than 10 TeV should be 
the CR interaction with ISM. The spectrum of $\gamma$-ray production is
calculated using a formalism by Dermer \citep{1986A&A...157..223D}
\begin{equation}
F(\epsilon_\gamma)=\int\limits_{\epsilon_\gamma+m_\pi^2/4\epsilon_\gamma}^{+\infty}
dE_\pi\frac{f(E_\pi)}{(E_\pi^2-m_\pi^2)^{1/2}}.
\end{equation}
Here $E_\pi$ is the total energy of the neutral pion and $m_\pi$ is its 
mass, $f(E_\pi)$ is the spectrum of neutral pions which is
\begin{equation}
f(p)=4\pi n_{\rm gas}\int dp'\frac{d\sigma(p,p')}{dp}n(p'),
\end{equation}
where $n_{\rm gas}$ is the gas density, $d\sigma(p,p')/dp$ is the 
production cross section, $n(p')$ is the nucleon density, $p$ and $p'$ are
the momenta of the pions and incident nucleons. The key points are the 
density distribution of CRs and the ISM distribution in the Galaxy. 

We extend the propagation scenario of the Galactic CRs described by 
GALPROP \citep{1998ApJ...509..212S} to the knee region. The spatial
and energy distribution of CRs is described by the solution of the
propagation equation
\begin{eqnarray}
\frac{\partial \psi(\vec{r},p,t)}{\partial t}& = &q(\vec{r},p,t)+
\nabla\cdot(D_{xx}\psi-V_c\psi) \nonumber\\
    &+& \frac{\partial}{\partial p}p^2D_{pp}\frac{\partial}{\partial p}
    \frac{1}{p^2}\psi-\frac{\partial}{\partial p}\left[ \dot p\psi -
    \frac{p}{3}(\nabla\cdot V_c\psi)\right] \nonumber\\
    &-&\frac{\psi}{\tau_f}-\frac{\psi}{\tau _r}
\end{eqnarray}
where $\psi(\vec{r},p,t)$ is density of CR particles per unit momentum 
$p$ at position $\vec{r}$, $q(\vec{r},p,t)$ is the source term, $D_{xx}$ 
is the spatial diffusion coefficient, $V_c$ is the convection velocity, 
$D_{pp}$ is the diffusion coefficient in momentum space and used to
describe the reacceleration process, $\dot p\equiv dp/dt$ is momentum 
loss rate, $\tau_f$ and $\tau _r$ are time scales for fragmentation and 
radioactive decay respectively. The spatial diffusion coefficient is
assumed to be space-independent and has a power law form 
$D_{xx}$ = $\beta D_0(\rho/\rho_0)^\delta$ of the rigidity $\rho$, 
where $\delta$ reflects the property of the ISM turbulence. 
The reacceleration can be described by the diffusion in momentum space 
and the momentum diffusion coefficient $D_{pp}$ is coupled with the spatial 
diffusion coefficient $D_{xx}$ as \citep{1994ApJ...431..705S}
\begin{equation}
D_{pp}D_{xx} = \frac{4p^2v_A^2}{3\delta(4-\delta ^2)(4-\delta)w} 
\end{equation}
here $v_A$ is the Alfven speed, $w$ is the ratio of magnetohydrodynamic 
wave energy density to the magnetic field energy density, which can be 
fixed to 1. The CRs propagate in an extended halo with a characteristic
height $z_h$, beyond which free escape of CRs is assumed.

In this work, we adopt the nominal propagation parameters of the diffusion
reacceleration scenario which are adjusted to reproduce the CR data such
as B/C, $^{10}$Be/$^9$Be, the local proton and electron spectra 
\citep{2010ApJ...720....9Z}. The major parameter values are 
$D_0=5.5\times10^{28}$ cm$^2$s$^{-1}$, $\delta=0.34$, $v_A=32$ km s$^{-1}$,
$z_h=4$ kpc.

The ISM is composed of about 10$\%$-15$\%$ of the total mass of the 
Galactic disk and its chemical composition is dominated by hydrogen and 
helium. The helium fraction of the gas is taken as 0.11 by number.
The hydrogen gas density $n_H$ includes three main components: 
molecular (H$_2$), atomic (H$_I$) and ionized (H$_{II}$). We use the gas distribution
in GALPROP, which is based on the survey results and related modeling
\citep{1988ApJ...324..248B,1976ApJ...208..346G,1991Natur.354..121C}.

\subsection{Diffuse $\gamma$-ray Emission}

\begin{figure*}[!htb]
\centering
\includegraphics[width=0.48\textwidth]{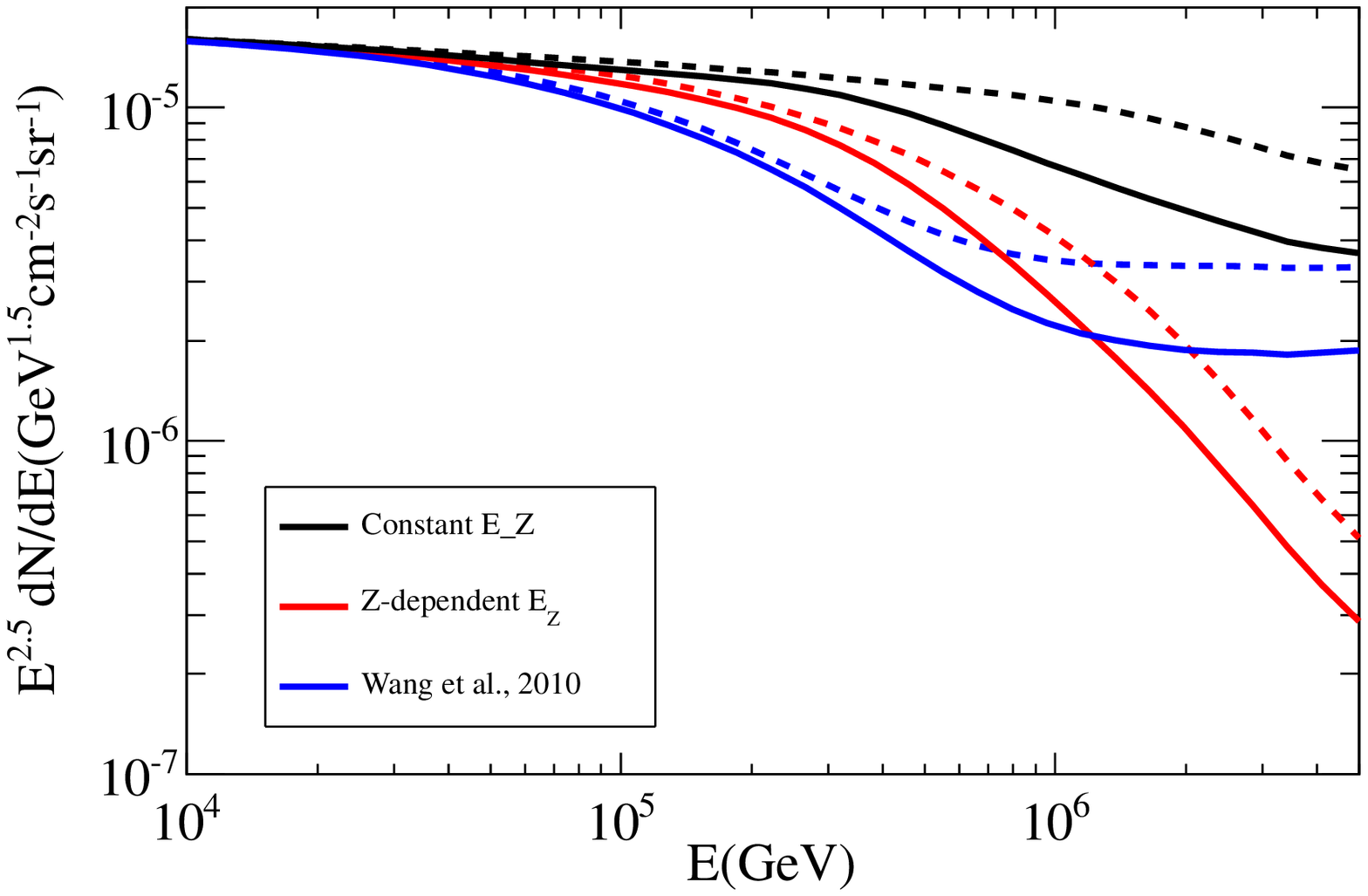}
\includegraphics[width=0.48\textwidth]{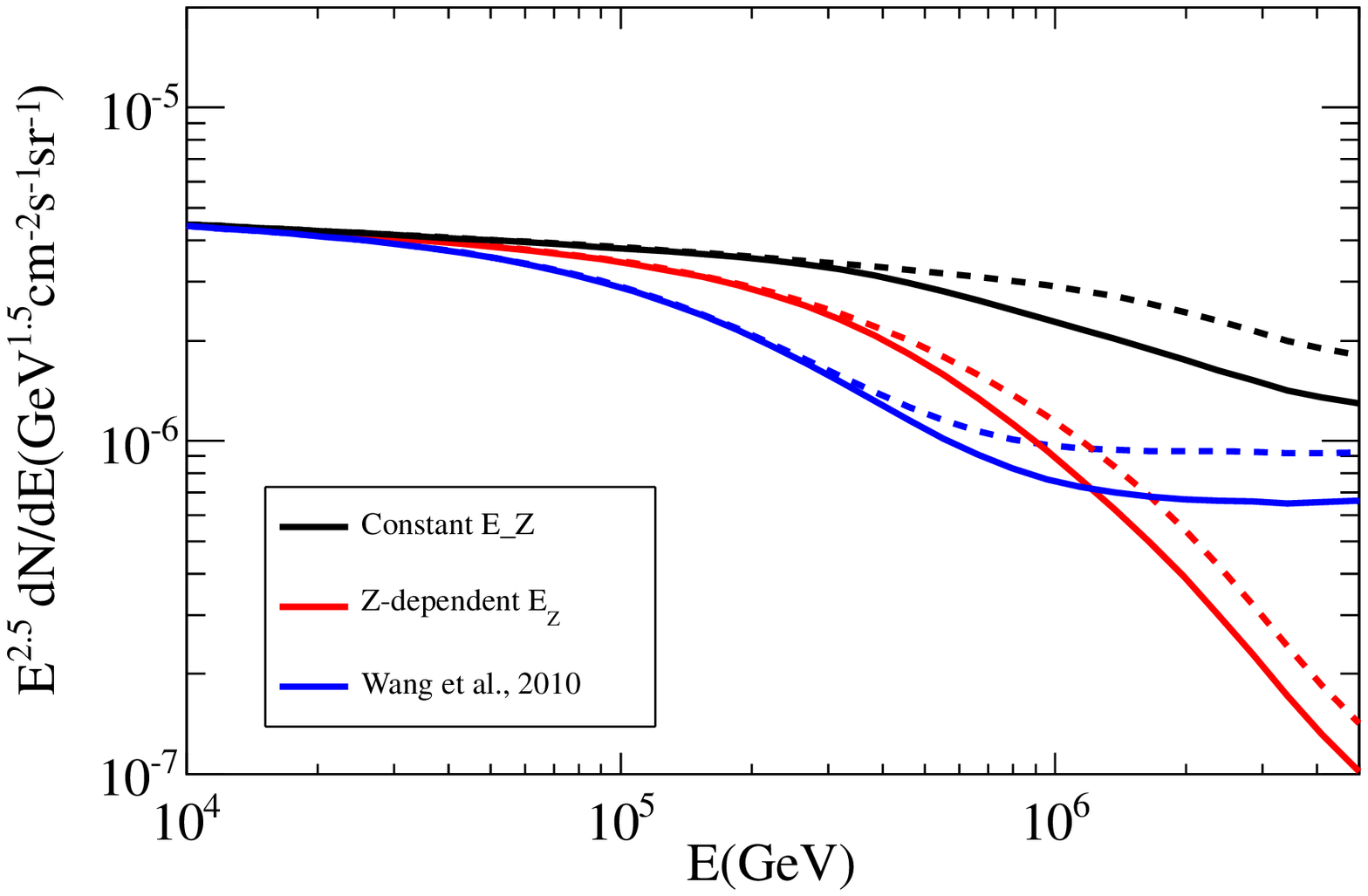}
\caption{Calculated $\gamma$-ray spectrum for inner ($20^{\circ}<l<55^{\circ}$,
$|b|<2^{\circ}$ ,left) and outer ($140^{\circ}<l<225^{\circ}$, $|b|<2^{\circ}$, 
right) Galactic plane regions. The dashed lines are the unattenuated spectra and 
the solid lines are the attenuated ones. Different colors show results for three 
models to describe the knee. See the text for details.}
\label{fig:gamma}
\end{figure*}

\begin{figure*}[!htb]
\centering
\includegraphics[width=0.48\textwidth]{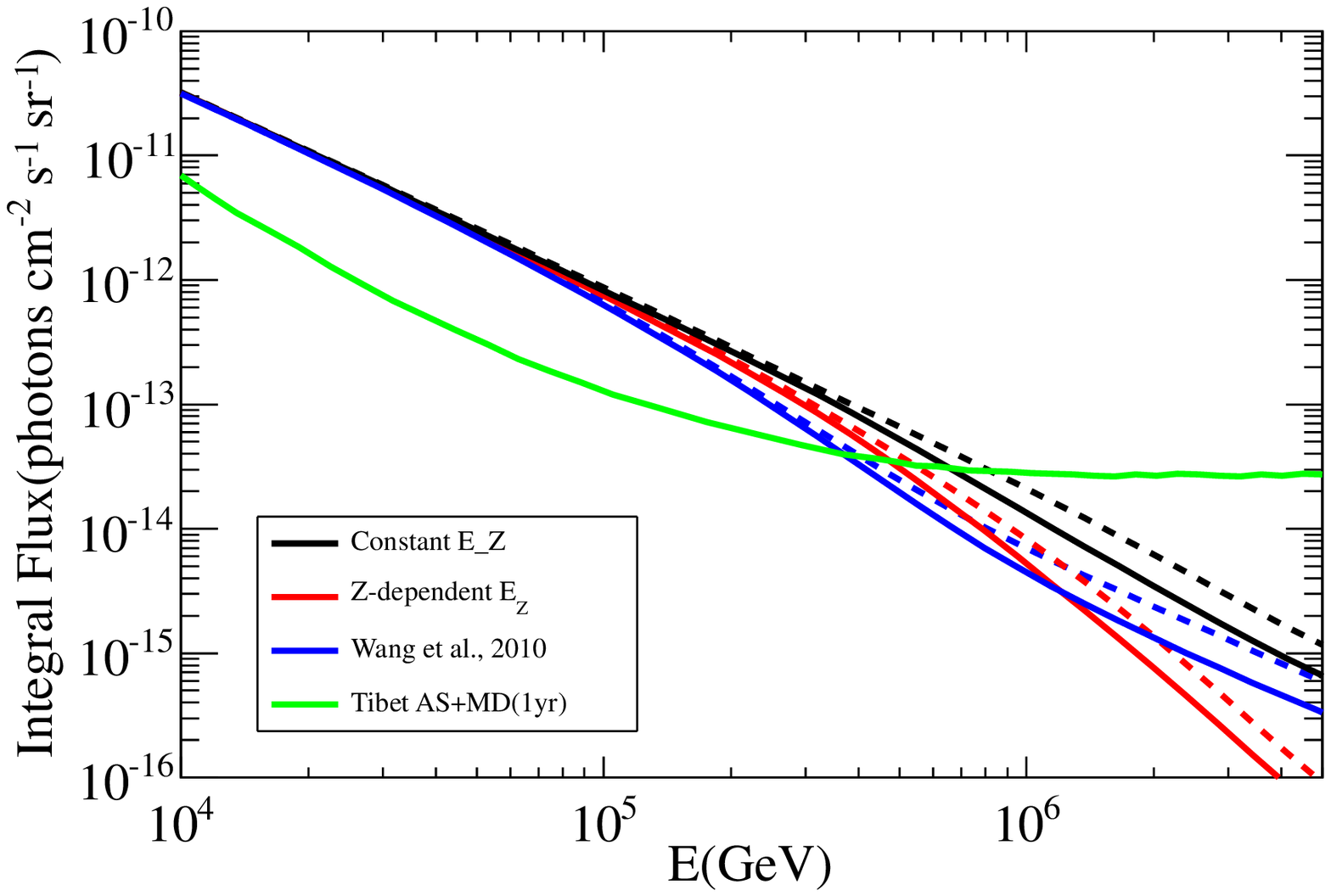}
\includegraphics[width=0.48\textwidth]{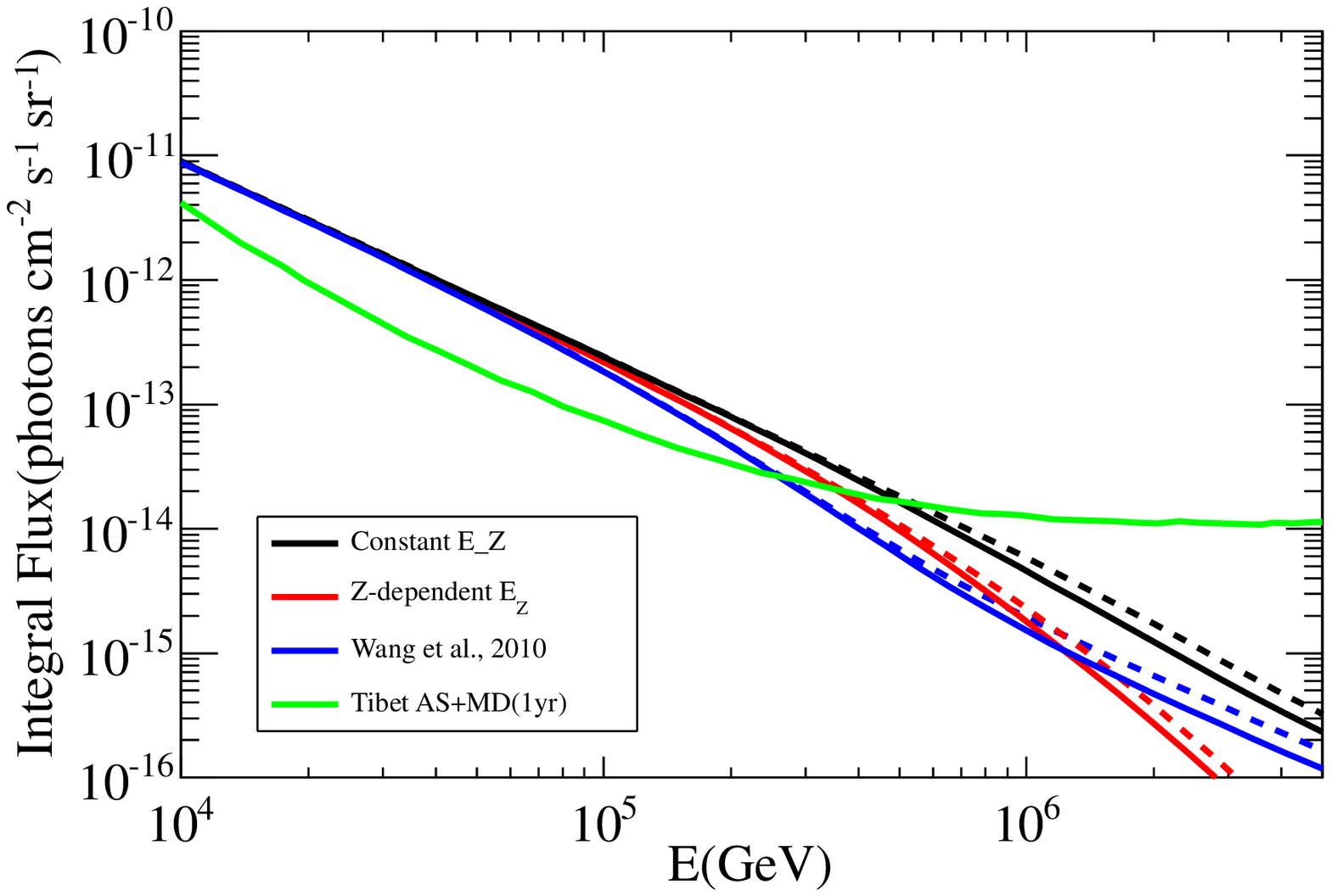}
\caption{Same as Figure \ref{fig:gamma}, but for integral spectrum, compared 
with the sensitivity of Tibet-AS$\gamma$+MD.} 
\label{fig:gamma2}
\end{figure*}

The spectrum of diffuse $\gamma$-ray emission can be calculated based on 
the calculated CR distribution in the Milky Way. In our calculation,
  the spectral index of CRs after propagation is $\sim$2.7 and the corresponding spectral index 
  of $\gamma$-ray is $\sim$2.6. Because the majority of 
the EAS experiments are located in the northern hemisphere, we choose an 
inner Galactic plane region ($20^{\circ}<l<55^{\circ}$ and $|b|<2^{\circ}$) 
and an outer Galactic plane region ($140^{\circ}<l<225^{\circ}$ and 
$|b|<2^{\circ}$) to display the results. 

The Galaxy is not transparent to very-high-energy $\gamma$-rays. The 
main three processes resulting in energy losses of photons are photoelectric 
effect, Compton scattering and pair production. The photoelectric effect is 
negligible for the very-high-energy $\gamma$-ray photons discussed here,
whose energies are higher than tens of TeV. As for the comparison between 
Compton scattering ($\gamma e$) and pair production ($\gamma\gamma$), we can 
compare the electron density with the ISRF target photon number density. 
The typical value of the interstellar gas density in the Galactic plane is 
$\sim1$ cm$^{-3}$. It means that the electron density is about 
1 cm$^{-3}$, which is much less than that of the infrared and microwave 
background photons \citep{2000ApJ...537..763S,2005ICRC....4...77P}. Therefore 
the Compton scattering losses of the $\gamma$-rays can also be neglected.
The dominant contribution to the attenuation of the-very-high energy 
$\gamma$-rays comes from the pair production. The optical depth as a function 
of photon energy and direction, $\tau(E,\psi)$, can be calculated as 
\citep{2006A&A...449..641Z,2006ApJ...640L.155M}
\begin{eqnarray}
\tau(E,\psi) & = & \int_{l.o.s.} dL \int d\cos(\theta) \int 
    \frac{dn(\epsilon,R,z)}{d\epsilon} \nonumber\\
    &\times& \sigma_{\gamma\gamma}(E,\epsilon,\cos\theta)
    \frac{1-\cos\theta}{2} d \epsilon,
\end{eqnarray}
where $\epsilon$ is the energy of the ISRF photon, $dn(\epsilon,R,z)/d\epsilon$ 
is the differential number density of ISRF which depends on the spatial location 
in the Galaxy, $\sigma_{\gamma\gamma}$ is the pair production cross section. 
The integral of $dL$ is along the line of sight (l.o.s.) of the incoming 
$\gamma$-ray photon. In our calculation, we adopt the ISRF model developed in 
\citet{2005ICRC....4...77P}, which is based on the new modelings of star and dust 
distributions, the scattering, absorption and re-emission of the stellar light 
by the dust.

The calculated diffuse $\gamma$-ray spectra are shown in Figure 
\ref{fig:gamma}. The left panel is for the inner Galactic plane region 
and the right panel is for the outer Galactic plane region. Different 
colors show the results of three different composition models of the 
knee: $Z$-dependent break (red), constant break
(black) and the physical interaction model (blue). The result of 
the $A$-dependent break model is similar to the $Z$-dependent one and is
not shown here. The dashed lines represent the unattenuated spectra and 
the solid lines are attenuated ones. 
It is shown that the $\gamma$-ray spectrum will also experience 
a knee-like structure around hundreds of TeV. The $Z$-dependent model 
predicts a big drop of the $\gamma$-ray spectrum, while the drop behavior 
of the constant break model is more shallow. For the interaction model,
a high-energy tail is left and forms a zigzag-like shape. However, the
attenuation effect makes the spectral behaviors more degenerate among
these models. To distinguish these models, we may need the precise 
measurements of the diffuse $\gamma$-ray spectrum up to energies beyond
several PeV.

To roughly investigate the detectability of such kinds of structures,
we compare the sensitivity of the Tibet-AS$\gamma$ experiment with muon
detectors (Tibet-AS$\gamma$+MD) \citep{2009APh....32..177S}. 
The sensitivity of the Tibet-AS$\gamma$ experiment is defined as a 5$\sigma$
observation of $\gamma$-ray excess over the survived CR background after muon detector rejection
    in one year of operation. 
Below 1 PeV, the CR background is rejected partly and the significance can be simply defined as
$N_\gamma/\sqrt{N_{CRs}}$. Above 1 PeV, the number of CR background is fully
suppressed down to less than one event per year and the sensitivity is defined as 10 $\gamma$-ray 
events.
Figure \ref{fig:gamma2} shows the integral flux and a comparison with the sensitivity of 
the Tibet-AS$\gamma$ experiment. 
The line labels are similar
to Figure \ref{fig:gamma}. With a few years of observations, Tibet-AS$\gamma$+MD 
may detect the diffuse $\gamma$-ray component up to several hundred TeV,
and the knee-like structure. However, it may still have difficulty  
discriminating different models, because the largest difference comes out
in PeV energies. The new generation high-energy $\gamma$-ray projects such 
as LHAASO \citep{2010ChPhC..34..249C} and HiSCORE \citep{2012AIPC.1505..821T} 
will have better capabilities to measure the $\gamma$-ray shape precisely 
and to distinguish different models. It is expected that LHAASO will 
have about a 30 times larger effective area than Tibet-AS$\gamma$+MD. If the 
background rejection of CRs is effective enough, LHAASO may be $\sim30$ 
times more sensitive than the current Tibet-AS$\gamma$+MD experiment.
In that case, we may expect that $\sim$year exposure could be very promising
to test the different models.
However, considering the Poisson fluctuation of $\gamma$-ray events to limit the
	capability,
    the LHAASO project requires a longer time of operation to discriminate the
   different line curve, such as five years with more than 50 $\gamma$-ray event observations.
 Furthermore, the possible systematics will make 
the case less optimistic, which depends on detailed simulation of the 
detectors \citep{2014APh....54...86C}.

\subsection{Diffuse Neutrino Emission}

The charged pion decay will produce neutrinos accompanied with the 
$\gamma$-rays. Different from $\gamma$-rays, neutrinos will propagate
freely in the space without absorption, which may carry
the information of the primary CRs more directly.

On average, the $pp$ collision produce $1/3$ neutral pions and $2/3$ 
charged pions. Each neutral pion decays into a pair of $\gamma$-rays, 
and each charged pion decays into two muon neutrinos and one electron 
neutrino (we do not distinguish neutrinos and anti-neutrinos). 
The initial neutrino flux ratio is approximately $\nu_e:\nu_\mu:\nu_\tau =
1:2:0$ from charged pion decay. The flavor ratio will be close to 
$\nu_e:\nu_\mu:\nu_\tau = 1:1:1$ at the Earth after vacuum oscillation 
with traversal of astrophysical distance. The typical energy of the 
neutrinos resulting from these decays is half of the $\gamma$-ray 
photons. Therefore the resulting neutrino spectrum is shifted relative 
to the source $\gamma$-ray spectrum. The typical spectrum of the 
muon neutrinos ($\nu + \bar {\nu }$) is then $2^{1-\Gamma}$ times
of the $\gamma$-ray spectrum, with $\Gamma$ the spectrum index of the
photon spectrum \citep{2006PhRvD..74f3007K}. 

\begin{figure}[!htb]
\centering
\includegraphics[width=0.48\textwidth]{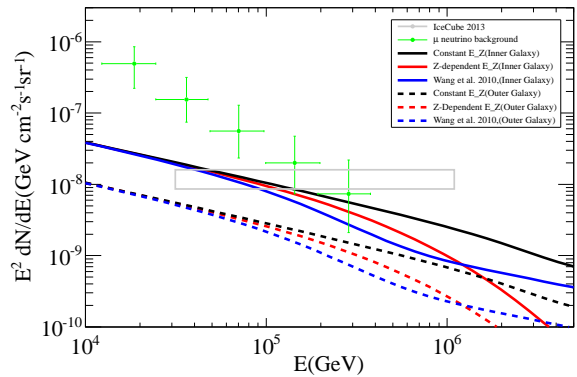}
\caption{Calculated diffuse neutrino spectrum from the collision of the
CRs with the ISM in the inner ($-30^{\circ}<l<30^{\circ}$, $|b|<5^{\circ}$, 
upper solid) and outer ($90^{\circ}<l<270^{\circ}$, $|b|<5^{\circ}$, lower 
dashed) Galactic plane regions. Also shown are the atmospheric muon neutrino 
background observed by IceCube \citep{2013PhRvL.111h1801A} and the recent 
IceCube results of possible astrophysical neutrinos \citep{2013PhRvL.111b1103A,
2013Sci...342E...1I}, adopted from \citep{2013PhRvD..88l1301M,
2014PhRvD..89h3004L}.}
\label{fig:neutrino}
\end{figure}

The calculated neutrino spectrum is shown in Figure \ref{fig:neutrino} 
for the inner Galactic plane ($-30^{\circ}<l<30^{\circ}$, $|b|<5^{\circ}$)
and outer Galactic plane ($90^{\circ}<l<270^{\circ}$, $|b|<5^{\circ}$).
Also shown is the estimated neutrino flux of recent IceCube observations
\citep{2013PhRvL.111b1103A,2013Sci...342E...1I}, adopted from
\citep{2013PhRvD..88l1301M,2014PhRvD..89h3004L}. In the inner region, the Galactic 
diffuse neutrinos are consistent with the IceCube observation in the $\sim100$ 
TeV energy region, but it is not enough to account for the high-energy
(up to PeV) events. Therefore it is possible that the Galactic neutrinos
from interactions between CRs and the ISM may contribute a proper fraction
to the low-energy events by IceCube \citep{2013PhRvD..88h1302R,
2013arXiv1309.4077A,2013APh....48...75G,2014MNRAS.439.3414J,2014arXiv1407.2536W}, 
while the rest events may require another origin with harder spectrum 
\citep{2013arXiv1307.1450H}. More and better measurements of the neutrino
events are necessary to clearly identify the Galactic component and
distinguish different models of the knee.

\section{Conclusion}

In this work, we propose an alternative method to pinpoint the CR 
compositions around the knee region with diffuse $\gamma$-rays and
neutrinos. It is shown that both $\gamma$-rays and neutrinos from
the interactions between the CRs and the ISM will experience a knee-like
structure at hundreds of TeV. Different models of the knee will predict
different behaviors of the generated $\gamma$-ray and neutrino spectra
due to the different compositions around PeV energies. Precise 
measurement of the diffuse $\gamma$-ray and neutrino spectra may be 
used to test different models to explain the knee. The newly upgraded 
Tibet-AS$\gamma$+MD experiment will have the potential to detect the
knee of the diffuse $\gamma$-rays. However, to test different models
of the knee we need future more sensitive experiments, such as LHAASO 
and HiSCORE. As a guaranteed source of neutrinos, we show that the
diffuse Galactic neutrinos may contribute to a proper fraction of the
recently reported neutrino events by IceCube at low energies
\citep{2013PhRvL.111b1103A,2013Sci...342E...1I}. The high-energy events,
however, require another origin with harder spectrum.

\section*{Acknowledgements}
This work is supported by the 973 program from the Ministry of Science and 
Technology of China (No. 2013CB837000), the Natural Sciences Foundation of 
China (Nos. 10725524, 10773011, 11135010, 11105155) and the Chinese Academy 
of Sciences (Nos. Y4546130U2). 

\bibliographystyle{apj}
\bibliography{knee}

\end{document}